\documentclass[twocolumn, showpacs,preprintnumbers,amsmath,amssymb]{revtex4}

\usepackage{graphicx}

\begin{document}

\addtolength{\textheight}{1.2cm}
\addtolength{\topmargin}{-0.5cm}

\newcommand{\etal} {{\it {\it et al.}}}

\title {Hysteresis loops of the magnetoconductance in graphene devices}

\author{A. Candini$^{1,3}$, C. Alvino$^{1,2}$, W. Wernsdorfer$^3$ and M. Affronte$^{1,2}$}

\affiliation{$^1$ S$^3$, Istituto Nanoscienze - CNR, via Campi 213/a, I-41125 Modena, Italy\\
$^2$Dipartimento di Fisica, Universit\`a di Modena e Reggio Emilia, via Campi 213/a, I-41125 Modena, Italy \\
$^3$ Institut N\'eel, CNRS, BP166, 25 Avenue des Martyrs, F-38042 Grenoble, France}


\begin{abstract}
We report very low-temperature magnetoconductance $\Delta G$ measurements on graphene devices with the
magnetic field $H$ applied parallel to the carbon sheet. The $\Delta G(H)$ signal depends on the gate voltage
$V_g$ and its sign is related to the universal conductance fluctuations. When the magnetic field is swept at
fast rates, $\Delta G$ displays hysteresis loops evident for different sizes and at different transport
regimes of the devices. We attribute this to the magnetization reversal of paramagnetic centers in the
graphene layer, which might originate from defects in our devices.
\end{abstract}


\maketitle

Among the multitude of fields of interest, carbon-based nano-materials are promising candidates for
applications in spintronics
owing to their low intrinsic spin-orbit effect, as well as the low hyperfine interaction of the electron
spins with the carbon nuclei. \cite{LossNatPhys, GeimRev} In this context, the transport properties of
graphene in the presence of a magnetic field have been deeply studied showing, for instance, that application
of a field perpendicular to the graphene plane induces quantum phenomena, such as the quantum Hall effect
\cite{KimNat, GeimNat}, or weak localization. \cite{MorozovPRL, SavchenkoPRL} Recently, magnetotransport
measurements have been used to characterize the transport regime in etched nano-devices,
\cite{Morpurgo_Moser_Bai} but relatively few studies have been carried out with the field applied along the
graphene layer, although these allow to study purely spin-related effects \cite{FolkNatPhys, EnsslinPRL}.

Several theoretical works have addressed the role of atomic scale defects, such as adatoms and vacancies, on
the local electronic structure of graphene and few-layers graphene, finding that they can actually carry a
magnetic moment. \cite{Palacios, NieminenPRL, NetoPRL} Also the possibility of long-range magnetic order has
been predicted for certain distributions of such defects. \cite{KatnelsonNatPhys, YazyevPRB} While evidence
of ferromagnetism has been reported for bulk graphite, \cite{EsquinaziPRL} the experimental results on
intrinsic magnetism in graphene are still controversial \cite{MagGraphene, Geim_Winpenny} also due to
difficult experimental conditions. Since the total magnetic moment of a single layer graphene is expected to
be extremely small, it has been suggested that a proof of magnetism in graphene can be accessed via
magnetotransport measurements. \cite{NetoPRB} However, none of the results present in the literature have
evidenced intrinsic magnetism in graphene through transport measurements so far.

Here we focus on low-field ($B < 1$T) parallel magnetoconductance (magnetic field along the graphene layer).
We show that when the field is swept at fast rates the magnetoconductance displays hysteresis loops that we
relate to the presence of magnetic impurities in the graphene layer.

We obtain graphene flakes by the standard mechanical exfoliation method from natural graphite. Thin flakes
are optically located with respect to pre-patterned alignment markers on top of $p$-doped silicon wafer
coated with 300 nm of oxide. The effective number of layers is checked by micro-Raman spectroscopy. Metal
contacts (Cr/Au or Ti/Pt) on the graphene sheets are obtained by electron beam lithography (EBL),
electron-beam or Joule evaporation and lift-off. The underlying doped silicon is contacted (from either the
top or the bottom) and is used as a back gate. By using this technique we routinely fabricate graphene
devices with carriers mobility $\mu$ up to 5,000 cm$^2$V$^{-1}$s$^{-1}$. Electrical measurements are
performed by using a lock-in amplifier with an applied ac voltage ($< 100 \mu$V at 33 Hz) in a dilution
fridge operated at a base temperature of 40 mK and equipped with a three-dimensional (3D) vector magnet with
sweeping rates as fast as 0.2 T/s.

\begin{figure}[ptb]
\begin{center}
\includegraphics[width=6.5cm]{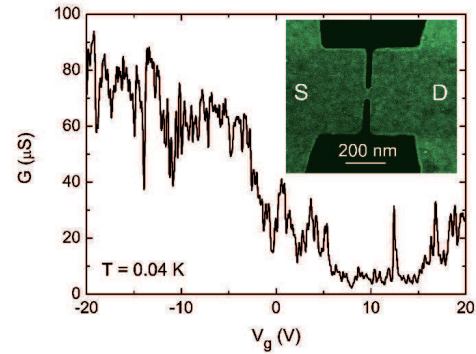}
\end{center}
\caption{(Color online) Source-drain conductance for varying back-gate voltage at 0.04K. Inset: False-color
scanning electron microscope picture of the nanoconstriction presented in the text. The graphene layer has
been colored in light green (gray) to enhance the contrast; the dark regions correspond to the substrate
after O$_2$ plasma etching of graphene. Source and drain part of graphene are indicated.} \label{fig1}
\end{figure}

For some of our devices, we employed a second step of EBL followed by low-energy oxygen plasma in order to
reduce the channel width. In this way we produce patterned ribbons, with different aspect ratios and lateral
sizes ranging from 500 nm down to 50 nm, and nanoconstrictions of different sizes ranging from 50 nm down to
10 nm. In the following we focus on results obtained with a constriction of size $\approx 60$ nm (see inset
of Fig. \ref{fig1}), for which the parallel magnetoconductance signal, defined as $\Delta
G=[G(B)-G(B=0)]/G(B=0)$, was stronger (patterned samples usually display larger magnetoconductance, up to
10\%, with respect to the nonpatterned ones, where the signal is usually below 1\%).

It is important to stress, however, that the main effect described in this work (hysteresis loop in the
magnetoconductance) is not related to a specific type of device, since it was observed in a variety of
different samples (examples are reported the supplemental material \cite{SupplInfo}). These include
non-patterned graphene sheets, with conductivity $G$ equal to a few $G_0$ ($G_0=2e^2/h$) and mobility up to a
few thousands cm$^2$V$^{-1}$s$^{-1}$ at 40 mK; nanoribbons of different sizes and aspect ratios with
conductivity $G$ between 0.1 and 1 $G_0$ and mobility (measured outside the transport gap region, when
present) $\mu \le 100$ cm$^2$V$^{-1}$s$^{-1}$; and small nanoconstrictions with suppressed low-temperature
conductivity ($G$ typically less than 0.1 $G_0$).

The low-temperature conductance $G$ of the nanoconstriction is shown in Fig. \ref{fig1} as a function of the
back-gate voltage $V_{\rm g}$. The curve is characterized by a flat low-conducting region around 10 V
(suggesting the incomplete formation of an electronic band gap) and by the presence of strong oscillations of
$\approx$ 0.1 $G_0$ in the signal. These oscillations, well reproducible as a function of $V_{\rm g}$,
evidence charge-carrier phase coherence phenomena (quantum interference) within our device.
\cite{Horsell_SSC}

\begin{figure}[ptb]
\begin{center}
\includegraphics[width=6.5cm]{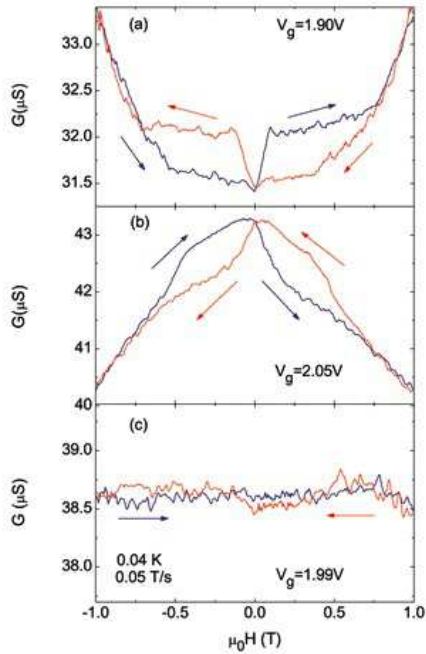}
\end{center}
\caption{(Color online) Parallel magnetoconductance of the devices at three different gate voltages. All
measurements have been performed at 0.04 K sweeping the magnetic field at a rate of 0.05 T/s. The curves have
been selected to show that the magnetoconductance can be positive (a), negative (b) or flat (c). In the (a)
and (b) cases, a hysteresis is observed.}\label{fig2}
\end{figure}

In Fig. \ref{fig2} we report the magnetoconductance curve as a function of an external field applied in the
plane of the graphene device, taken at three representative gate voltages $V_{\rm g}$. We notice that the
low-temperature parallel magnetoconductance can be either positive or negative on the same device, depending
on the value of $V_{\rm g}$. This effect is different from the behavior reported for magnetoconductance
measurements with magnetic field applied perpendicular to the graphene sheet, for which the sign is
predominantly found to be positive in small etched devices, \cite{Morpurgo_Moser_Bai} while in larger flakes
is dominated by weak localizations effects. \cite{SavchenkoPRL}

Remarkably, for low fields the $\Delta G(B)$ curve depends on the field sweep rate cycle and we observe the
opening of a hysteresis loop. This can be characterized by the difference $\delta G$ between the two curves
obtained with increasing and decreasing the field. We notice that (i) the sign of $\delta G$ depends on the
sign of the magnetoconductance $\Delta G$ and (ii) $\delta G$ changes sign with the magnetic-field
orientation. In this way $\delta G$ is negative for $B < 0$ and positive when $B
> 0$ in the case of positive magnetoconductance [Fig. \ref{fig2}(a)], and the opposite in the negative
magnetoconductance case (Fig. \ref{fig2}(b)). No hysteresis is detected when the magnetoconductance is flat
[Fig. \ref{fig2}(c)].

\begin{figure}[ptb]
\begin{center}
\includegraphics[width=6.5cm]{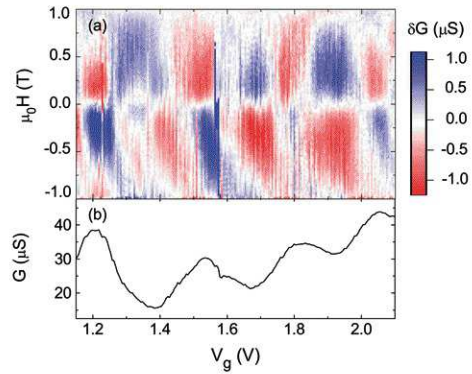}
\end{center}
\caption{(Color online). (a) Color scale plot of the hysteresis $\delta$$G$ (defined as the difference
between the curve recorded while upsweeping the magnetic field and the curve recorded while the field is
ramped back) vs gate voltage. (b) Corresponding zero-field conductance for the same gate region. $\delta$$G$
is maximum in correspondence of a maximum or a minimum in $G(V_{\rm g}$) and its sign is reversed between
them.} \label{fig3}
\end{figure}

To show the dependence of  $\delta G(V_{\rm g})$ on the back gate voltage in more detail, in Fig.
\ref{fig3}(a) we plot in color code the intensity of $\delta G$ as a function of $V_{\rm g}$ for a small
interval where several oscillations of the conductivity are present [Fig. \ref{fig3}(b)]. Clearly, $\delta
G(V_{\rm g})$ is maximum when $G(V_{\rm g})$ presents either a minimum or a maximum, while its sign with
respect to the field orientation is inverted between them. The corresponding magnetoconductance curves have
positive and negative sign and can be represented by the curves plotted in Figs. \ref{fig2}(a) and
\ref{fig2}(b), respectively. At the inflection point between a maximum and a minimum in $G(V_{\rm g})$ the
magnetoconductance is flat and the hysteresis is not observed: this situation corresponds to the case
depicted in Fig. \ref{fig2}(c).

\begin{figure}[ptb]
\begin{center}
\includegraphics[width=6.5cm]{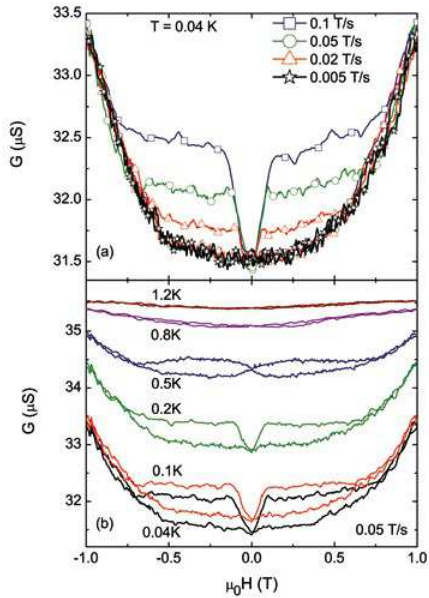}
\end{center}
\caption{(Color online) (a) Hysteresis loops obtained for different field sweeping rates at base temperature.
The loop is strongly rate dependent and it disappears for slow sweeping rates. (b) Evolution of the
magnetoconductance curves as a function of temperature sweeping the field at 0.05 T/s. The hysteresis loop is
still present up to $\sim 1$K, becoming undetectable when the magnetoconductance signal is flat and dominated
by noise.} \label{fig4}
\end{figure}

The hysteresis strongly depends on the magnetic-field sweeping rate, as evidenced in Fig. \ref{fig4}(a): for
fast rates, such as 0.1 T/s, it closes up for $B \sim 1$ T and its intensity is higher, while it smoothly
disappears as the field is swept slower. At a rate of 0.02 T/s the hysteresis closes at a field $B \sim 0.5$
T and $\delta G$ is about four times smaller than what is found for a rate of 0.1 T/s. The hysteresis becomes
practically undetectable when the field is swept slower than $\sim$0.01 T/s. This sets a time scale of few
tens of seconds for the appearance of the hysteresis loops. For the slowest sweep rates, the reversible,
i.e., non hysteretic, magnetoconductance signal is still present. Thus we may distinguish two distinct
components of the magnetoconductance: the equilibrium part of the $\Delta G$ signal, and the hysteretic part,
which depends on the sweeping rate. We notice that $\Delta G$ always shows the equilibrium values when the
field crosses zero (i.e., $\delta G=0$), while the hysteresis shows up only when the field is raised, either
with positive or negative direction, away from zero at relatively high sweeping rates. The behavior is that a
stronger magnetoconductance signal is observed just after passing zero field, relaxing toward the equilibrium
curve at higher fields [Fig. \ref{fig2}(a) and (b)].

The situation changes when the temperature raises, as shown in Fig. \ref{fig4}(b). As the temperature is
increased, the magnetoconductance becomes flat and the hysteresis disappears accordingly, becoming
practically undetectable above 1 K. Therefore at high temperatures the situation is similar to what is
observed for $V_{\rm g}$ fixed at an inflection point of $G(V_{\rm g}$) [Fig. \ref{fig2}(c)].

As previously mentioned, the hysteresis loop was reproducibly found in many different devices. In particular,
the appearance and the main features of the hysteresis loop (including temperature and field sweep rate
dependencies) do not depend on the size of the device (see supplemental information \cite{SupplInfo}). For
all the measured samples, no anisotropy is detected, as long as the field is applied in the plane of
graphene. In addition, we found that the behavior is quite insensitive to many post-processing procedures,
like annealing. Indeed, we systematically checked that the hysteresis loops do not substantially change after
annealing in Ar: we used temperatures from 200-400$^\circ$ C and times from 15 min to 2 h. Finally it is
worthwhile to stress that hysteresis loops were observed in different apparatus and that we carefully checked
our setup to exclude any possible experimental artifact.

To interpret these results, we first argue that the observation of a nonequilibrium phenomenon (hysteresis
loop of magnetoconductance) at the time scale of tens of seconds can be hardly explained within the framework
of transport models referring to solely steady state quantities. Indeed, the characteristic time scales
associated to scattering mechanisms in graphene are usually found to be orders of magnitude smaller
\cite{SavchenkoPRL}. The presence of electron-hole puddles domains, that are known to move under a magnetic
field, \cite{Broto_NJP_ElecHolePud} cannot play an important role here, since the hysteresis loop has been
found in different transport regimes, and even at relatively high doping level (see Fig. S1 of the
supplementary information \cite{SupplInfo}), where the puddles are expected to play a marginal role.
\cite{Yacoby_NatPhys}

On the other hand, the observed behavior reminds the processes of magnetization reversal of small magnetic
moments at low temperature. \cite{WW_V15} Based on these observations, we model our device to consist of two
physical parts: an electrical conductor (charge carriers in the graphene layer) that is coupled to localized
magnetic moments, as described, for instance, in Ref. \cite{NetoPRB}. In this simple model, the hysteresis
loops reflect the magnetization reversal of the localized moments, as the conducting graphene layer detects
the magnetization behavior through its magnetoconductance.

We first address the origin of the parallel magnetoconductance in the conducting layer and its change of sign
with the gate voltage $V_{\rm g}$. In the nonballistic regime, magnetoconductance arises from quantum
interference effects. \cite{efros} Results in Figs. \ref{fig2} and \ref{fig3} show that the
magnetoconductance is positive when $G(V_{\rm g}$) has a minimum, negative when $G(V_{\rm g}$) has a maximum,
and it is flat when $G(V_{\rm g}$) has an inflection point. In addition, in Fig. \ref{fig4}(b) it is shown
that the intensity of the magnetoconductance decreases as the temperature is raised, which is the same trend
that we observed for the oscillations in the conductance $G(V_{\rm g})$ signal. Thus the in-plane magnetic
field tends to suppress the fluctuations in the conductance, an effect already observed in two-dimensional
electron gases in heterostructures of semiconductors, \cite{FolkPRL, DebrayPRL},but this is the first time,
to our knowledge, that it is reported for graphene. Since orbital effects should be excluded in our case, we
tentatively relate this effect to the spin split of the density of electronic states at the Fermi level of
the conducting layer. More details depend on the specific regime considered as, for instance, discussed in
Refs.\cite{DasSarmaPRB} or \cite{NetoPRB}, but they will bring the discussion beyond of the scope of this
work.

The appearance of local magnetic moments in graphene has been widely studied theoretically \cite{Palacios,
NetoPRB, Harigaya} and, more recently, also experimentally, \cite{MagGraphene, Geim_Winpenny} and it was
ascribed to the presence of defects. These may include vacancies, \cite{Palacios} metals or H atoms.
\cite{NieminenPRL, NetoPRL, NetoPRB} Magnetic moments are also expected to be present at the edges of
graphene. \cite{Harigaya} Concerning our devices, since the observed behavior is essentially insensitive to
annealing process, it is unlikely that such impurities are dominated by H adatoms. \cite{GeimScience} We also
exclude an important role from edge defects, since the hysteresis is observed also in nonetched graphene
layers and changes only very little with the patterning process. Concerning the presence of metal impurities,
while we can not completely exclude the possibility of unwanted contaminations, a series of experimental
facts make us confident in excluding this hypothesis as the dominant source of the magnetic signal. Indeed we
did not observe any change when employing different solvents or solvents from different batches. In addition,
we prepared our samples employing two independent procedures to clean the SiO$_2$ substrate before the
deposition of graphene, namely O$_2$ plasma and Piranha cleaning, without finding variations in the signal.
Also a possible role of the electrical contacts has to be excluded, since, as already mentioned in the first
part of this Rapid Communication, different metals have been employed without finding any change in the
signal. Therefore we are led to conclude that structurally intrinsic defects, \cite{Geim_Winpenny} such as
vacancies, provide the dominant contribution to the formation of local magnetic moments. In Ref.
\cite{NetoPRB}, Rappoport and co-workers calculated the magnetoresistance of a graphene in the presence of
magnetic impurities, considering the case of a specific type of magnetic moments (H adatoms) in the variable
range hopping regime. Although the behavior we observed is more general, we compare their prediction on the
magnetoresistance to estimate the density of the magnetic impurities. According to their Eq. (6), we find
that the difference between the magnetoresistance at $B=0$ and high field is related to the total
magnetization $M_{\rm S}$ of the impurities, and hence on their total number. Comparing their predictions of
Fig.7 (where they found a magnetoresistance drop of 60\%) with our results (changing from $1\%$ to $3\%$ for
the highest sweep rates), we found that in our case the impurity concentration is therefore $\approx$ 100 -
200 times smaller, giving an approximate concentration of 250-500 ppm. Interestingly, this rough estimation
is in the same order of magnitude as that found in Ref. \cite{Geim_Winpenny}.

We now turn our attention to the origin of the magnetic hysteresis. Localized magnetic moments induced by
defects in graphene are not expected to possess magnetic anisotropy, due to the very small spin-orbit
interaction. Yet, the presence of ripples can also increase spin-orbit coupling, giving rise to a sizable
magnetic anisotropy. In addition, it has been theoretically shown that localized spins may interact among
them through Ruderman-Kittel-Kasuya-Yosida interaction mediated by free electrons, \cite{NetoPRB} and
eventually give rise to ordered magnetic structures. \cite{KatnelsonNatPhys, YazyevPRB}

At very low temperatures magnetic moments may lack phonons to relax with, maintaining their magnetization
even when the external field is reversed. This process, giving rise to a typical butterfly-shaped hysteresis
loop and happening at the time scale of seconds, becomes visible at a field sweep rate of a few tens of mT/s
and rapidly disappears for longer time scale. The phenomenon is known as phonon bottleneck and it has been
studied for small spin systems. \cite{WW_V15} For $S = 1/2$ impurities, as predicted by the majority of
theories on individual vacancies, we expect the hysteresis to rapidly disappear as the temperature is
increased or a transverse magnetic field is applied. However, Fig. \ref{fig4}(b) shows that the
magnetoconductance hysteresis of our graphene devices survives at temperatures as high as 800 mK. In
addition, we found that a transverse field larger than 0.5 T is necessary to completely suppress the
hysteresis. Therefore we conclude that the localized magnetic moments should have $S \geq 1/2$, consistent
with the phonon-bottleneck mechanism. This is also consistent with what was recently found by superconducting
quantum interference device measurements. \cite{Geim_Winpenny}

In conclusion, we observed hysteresis loops in the magnetoconductance of graphene devices with magnetic field
applied parallel to the conducting layer and at very low temperature ($T<1$ K). The associated time scale is
of the order of a few tens of seconds. The observed hysteresis loops are evident for different sizes and at
different transport regimes of our devices. We ascribe this behavior to the magnetization reversal of
localized spin moments, probably arising from defects in graphene. The magnetization reversal  is detected
via magnetotransport properties of the charge carriers within the graphene layer. Based on the field and
temperature dependence of the hysteresis, we conclude that the spin of the localized moments is higher than
$S=1/2$, in agreement with recent works. We believe that the design and the results of our experiments
constitute the basis for the functioning of more complex graphene-based spintronic devices.

\section {Acknowledgments}

This work has been supported by FP7-ICT FET Open "MolSpinQIP" project funded by EU, Contract No. 211284,the
ERC advanced grant MolNanoSpin (Grant No. 226558), and the ANR-Pnano project MolNanoSpin. The authors thank
V. Reita, E. Eyraud, L. del-Rey, D. Lepoittevin, R. Haettel, and the Nanofab facility for technical support.

\end{document}